\documentclass[twocolumn,showpacs,amsmath,amssymb,aps,prb,superscriptaddress,floatfix]{revtex4}
\usepackage{graphicx}
\usepackage{natbib}
\def\bea{\begin{eqnarray}}
\def\eea{\end{eqnarray}}
\def\be{\begin{equation}}
\def\ee{\end{equation}}  
\def\S{\mbox{\bf S}}
\def\et{{\it et al.}}
\begin{document}
\author{Andreas L\"auchli}
\email{laeuchli@comp-phys.org}
\affiliation{Institut Romand de Recherche Num\'erique en Physique des Mat\'eriaux (IRRMA),
  PPH-Ecublens, CH-1015 Lausanne}
\author{Guido Schmid}
\affiliation{Theoretische Physik, ETH H\"onggerberg, CH-8093 Z\"urich, Switzerland}
\altaffiliation[Present Address: ]{Supercomputing Systems AG, Technoparkstrasse 1, CH-8005 Z\"urich, Switzerland}
\author{Simon Trebst}
\affiliation{Microsoft Research and Kavli Institute for Theoretical Physics, University of California, Santa Barbara, CA 93106, USA}
\date{\today}
\title{Spin nematic correlations in bilinear-biquadratic $S=1$ spin chains}
\begin{abstract}
We present an extensive numerical study of spin quadrupolar correlations in single and coupled
bilinear-biquadratic spin one chains, using several methods such as Exact Diagonalization, Density
Matrix Renormalization Group and strong coupling series expansions.
For the single chain we clarify the dominant correlation function in the enigmatic gapless period-three
phase for $\theta \in (\pi/4,\pi/2)$, which is of spin quadrupolar nature with a period three spatial structure.
Then we revisit the open problem of the possible existence of a ferroquadrupolar phase between the
dimerized and the ferromagnetic phases. Although an extended critical region is in principle compatible with
the numerical results, a scenario with a huge crossover scale is more plausible.
Finally we study the fate of the dimerized phase upon coupling two chains in a ladder geometry. The
dimerized phase rapidly vanishes and an extended gapped phase takes over. This gapped phase 
presumably has dominant short-ranged ferroquadrupolar correlations for $\theta \in (-3\pi,4,-\pi/2)$
and -- suprisingly -- seems to be adiabatically connected to the plaquette single solid phase of the
Heisenberg $S=1$ ladder and therefore also with the Haldane phase of isolated chains.
  \end{abstract}
\pacs{75.10.Jm, 75.10.Pq, 75.40.Mg, 75.40.Cx}
\maketitle

\section{Introduction}

\begin{figure}[b]
  \centerline{\includegraphics[width=0.7\linewidth]{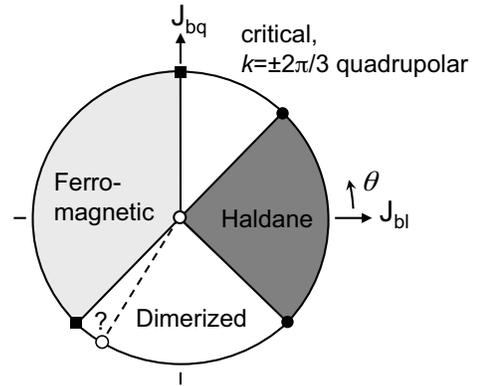}}
  \caption{
    Phase diagram of the bilinear-biquadratic spin-one chain.
    The firmly established phases are the Haldane, the ferromagnetic and the 
    dimerized phase.
    We characterize the extended gapless phase $\pi/4 \le \theta < \pi/2$ by having 
    dominant $k=\pm2\pi/3$ spin quadrupolar correlations.
    The possible occurence of a spin nematic like phase close to $-3\pi/4$ is investigated
    and critically discussed. 
    \label{fig:PhaseDiagram}}
\end{figure}

The recent experimental demonstration \cite{OrzelGreiner} of the transition from 
a superfluid state to a Mott insulating state of atoms in an optical lattice has opened the
way to novel realizations of effective quantum lattice models with widely tunable control
parameters. Quantum magnetic systems can be realized by spinor atoms in an optical lattice, e.g. 
$^{23}$Na with a total $S=1$ moment. Confining $S=1$ atoms to an optical lattice there
are two scattering channels for identical atoms with total spin $S=0,2$ which can be 
mapped to an effective bilinear and biquadratic spin interaction \cite{Yip,Demler,Garcia-Ripoll}:
\be
H=\sum_{\langle i,j \rangle} \left[ J_{bl} \left(\S_{i}\cdot\S_{j} \right) 
  + J_{bq} \ \left(\S_{i} \cdot\S_{j} \right)^2\right],
\label{eqn:biquadHamiltonian}
\ee
where we adopt the standard parametrization $J_{bl}=\cos \theta$ and $J_{bq}=\sin\theta$. 
In one dimension, the bilinear-biquadratic spin-one model has a rich phase diagram
(see Fig.~\ref{fig:PhaseDiagram}) with some well established phases: 
the Haldane gap phase~\cite{HaldaneConjecture}, a dimerized phase \cite{DimerizedPhase}, 
a ferromagnetic phase and some less well understood phases: a critical phase with
period three correlations \cite{FathPeriodTripling}, which we will characterize as
having dominant spin nematic correlations, and possibly a gapped spin
nematic phase between the dimerized and ferromagnetic phase \cite{ChubukovPRB}, which
however remains controversial.

On the square and the simple cubic lattice, the bilinear-biquadratic spin-one model is well 
understood for the case $J_{bq}\le 0$. It exhibits a ferroquadrupolar spin nematic phase 
for $\theta \in (-3\pi/4,-\pi/2)$ \ \cite{PapaUnusualPhases,HaradaKawashimaPRB,TakanoExact}.
Adjacent to it are an antiferromagnetic N\'eel phase and a ferromagnetic phase 
\cite{PapaUnusualPhases,HaradaKawashimaPRB}. 
The region of purely antiferromagnetic couplings $J_{bl}, J_{bq}>0$ remains to be understood,
and could possibly contain a spin liquid region~\cite{SUNliquids}. Recently the Hamiltonian~(\ref{eqn:biquadHamiltonian})
on the triangular lattice attracted some interest \cite{TsunetsuguArikawa,AMP, BhattacharjeeSenthil06},
as a possible explanation of the unconventional magnetism of NiGa$_2$S$_4$ \cite{Nakatsuji}.

For a single chain Chubukov~\cite{ChubukovPRB} suggested the existence of a gapped, 
nondimerized phase showing dominant spin nematic correlations close to the ferromagnetic 
region of the phase diagram.
Subsequent numerical work~\cite{FathNematic} could however not substantiate this claim and 
it was therefore believed for a while that the dimerized phase would extend up to the 
ferromagnetic phase boundary. Recent quantum Monte Carlo calculations \cite{KawashimaPTP} and
field theoretical work~\cite{IvanovKolezhuk} suggest that this picture might need to be reconsidered,
especially in the light of possible experimental verifications in Bose-Einstein condensate 
systems~\cite{Yip,Garcia-Ripoll}. In the meantime there has been a considerable number of publications
in favor or against a spin nematic phase close to $\theta=-3\pi/4$\ \cite{LSTpreprint,Porras,Rizzi,Buchta,Bergkvist},
but leaving the final answer still open.

The aim of the present paper is twofold. First we give a characterization in terms of the dominant
correlation function for the extended gapless critical phase, whose mere existence is well established.
We find in this case that the predominant correlations are of spin quadrupolar nature, with a wavevector 
of $\pm 2\pi/3$.  The second aim is to shed some more light on the region $\theta \in [-3\pi/4,-\pi/2]$ of 
the single chain. We present extensive numerical simulation results based on different numerical methods 
and discuss possible interpretations. Due to the difficulty in obtaining a firm conclusion regarding the 
spin nematic phase, we have also studied bilinear-biquadratic ladders. In this case the dimerized phase 
plays only a minor role in the phase diagram and we report an extended gapped spin
nematic phase, very close in spirit to the proposal by Chubukov \cite{ChubukovPRB} for the single chain.
Surprisingly this phase seems to be adiabatically connected to the well known Haldane phase of the isolated
chain.

The outline of the paper is the following: in section~\ref{sec:period3} we show that the established 
gapless phase 
$\theta \in (+\pi/4,\pi/2)$ has dominating spin quadrupolar correlations,
a characterization which was lacking before.
We then move on to the region $\theta \in [-3\pi/4,-\pi/2]$ and discuss the possible 
existence of a spin nematic intermediate phase between the dimerized and the 
ferromagnetic phase. We report several anomalous physical properties encountered 
upon approaching $\theta \rightarrow -3\pi/4$, which could be interpreted as a phase 
transition to a spin nematic state.
However, due to numerical limitations it has remained elusive to pinpoint such a phase
transition, and an alternative scenario where a a very large crossover scale emerges as one approaches
the $SU(3)$ point at $-3\pi/4$ becomes more plausible.
In section~\ref{sec:twochains} we introduce an extension of the single chain model by 
coupling two chains to form a ladder. We show that the dimerized phase gives very rapidly 
way to an extended gapped phase. This short-ranged ordered phase encompasses the 
Haldane phase at $\theta=0$ as well as its ladder extension, the so called 
``Plaquette Singlet Solid'' state\cite{TodoLadder}, and crosses over to a gapped spin nematic 
state close to $\theta \rightarrow -3\pi/4$.

\section{The enigmatic period 3 phase}
\label{sec:period3}

The existence of an extended critical phase in the interval $\theta \in [+\pi/4,\pi/2)$
has first been discussed in Ref.~[\onlinecite{FathPeriodTripling}]. Later work
in Refs.~[\onlinecite{XiangTrimerized,ReedGapless,Bursill,ItoiKato,MuellerKarbach}]
agreed on the gapless nature of the phase due to soft modes at $k=0,\pm2\pi/3$.
The special point $\theta=+\pi/4$ has an enlarged  $SU(3)$ symmetry and is solvable
by Bethe ansatz \cite{Uimin,Lai,Sutherland}  (Uimin-Lai-Sutherland model), proving
the existence of soft modes at $k=0,\pm2\pi/3$ as a rigourous result.

Despite the consensus on the presence of an extended gapless phase, the nature of the 
dominant correlations in this phase has not been clearly worked out. 
Initally Xiang \cite{XiangTrimerized} proposed an almost ``trimerized'' groundstate, 
with dominant singlet correlations involving three consecutive spins. Although models 
can be constructed which have exactly trimerized groudstates \cite{SolyomExactTrimerized},
in the present model these are not the dominant fluctuations, as shown earlier in 
Ref.~[\onlinecite{MuellerKarbach}].

Here we show that the dominant correlations are not of singlet,
but of {\em spin nematic}, i.e. quadrupolar character.
We build upon the field theoretical work of Itoi and Kato \cite{ItoiKato} in 
which they showed that 
throughout the critical region there are three fields with spin $S=0,1,2$ whose 
scaling dimensions are all equal to $x=2/3$.
However apart from $\theta=\pi/4$ it is the $S=2$ mode at $k=\pm2\pi/3$ -- which is
subject to correlation-{\em enhancing} logarithmic corrections -- which will show
the dominant correlations. Note that this is consistent with the observation~\cite{FathPeriodTripling}
that the lowest energy level at $k=\pm2\pi/3$ on finite chains carries $S=2$.

\begin{figure}
  \centerline{\includegraphics[width=0.95\linewidth]{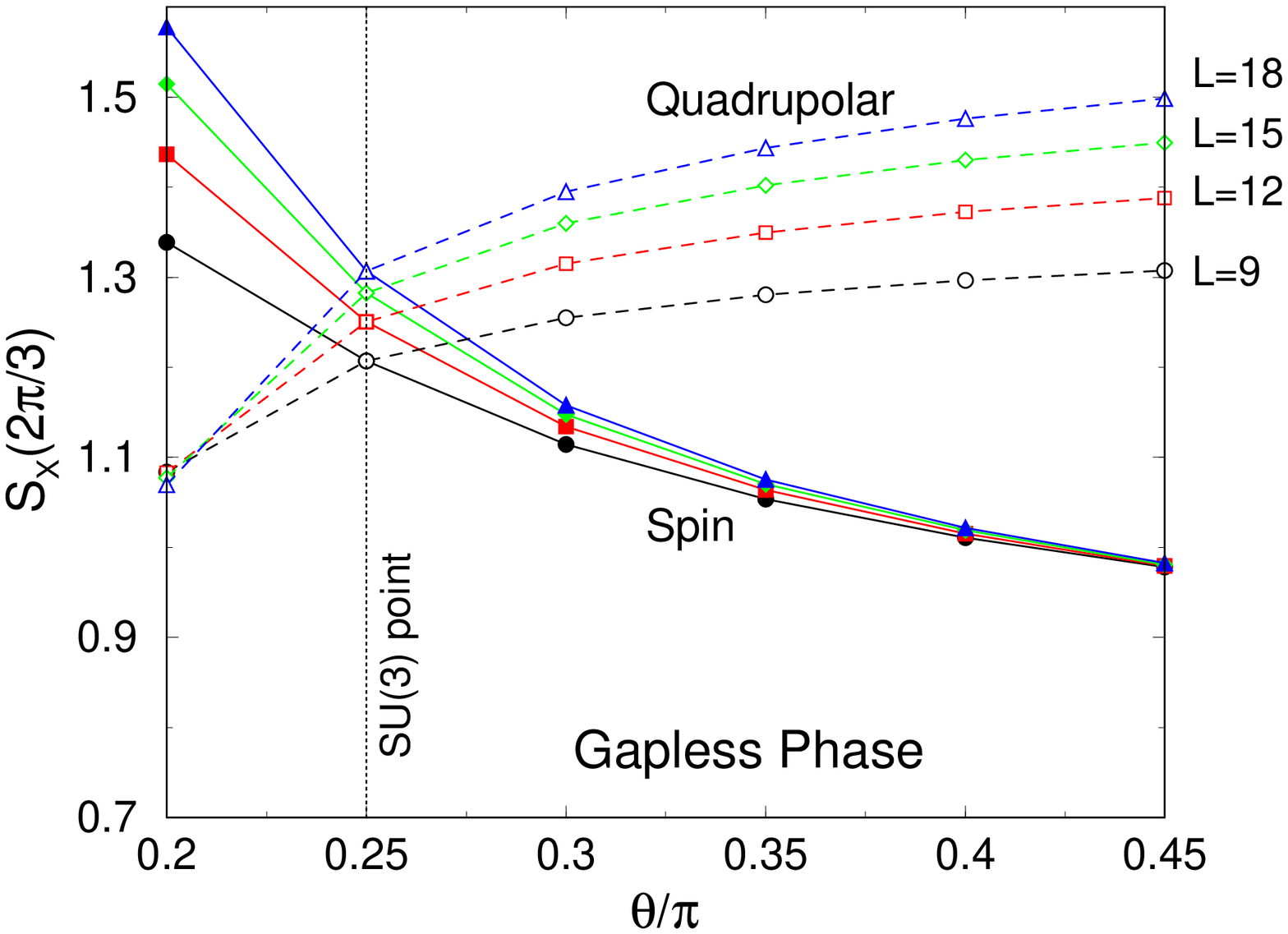}}
  \centerline{\includegraphics[width=0.84\linewidth]{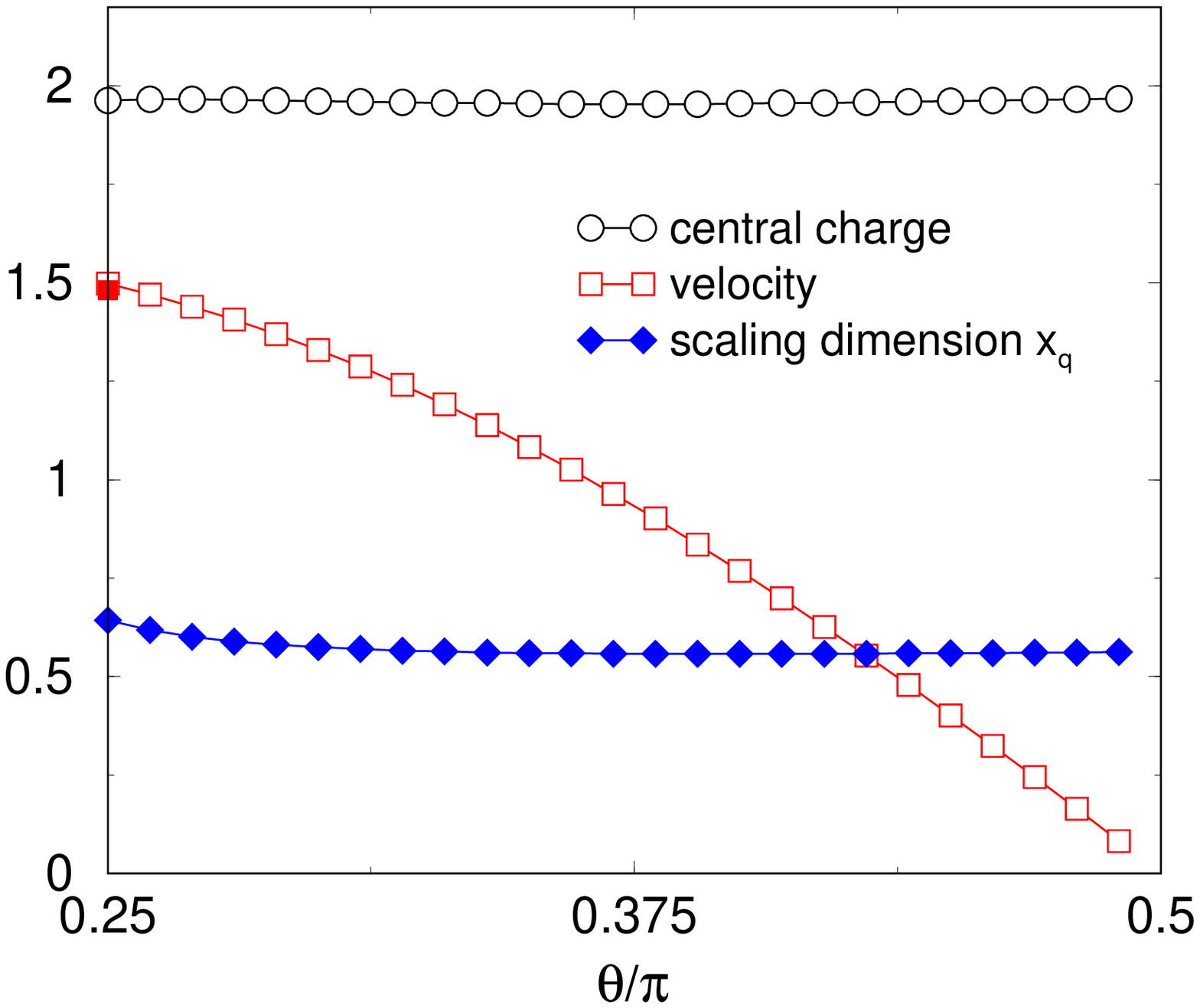}}
  \caption{
   (Color online)
    Upper plot:
    Static structure factors for spin and quadrupolar correlations
    at momentum $k=2\pi/3$ as a function of $\theta$ for different
    system sizes. At the SU(3) point $\theta=+\pi/4$ the two correlation functions
    are related by symmetry. By going deeper into the critical region the spin 
    correlations get weaker, while the quadrupolar correlations are enhanced. 
    Lower plot: the central charge, the excitation velocity and the scaling dimension of the
    quintuplet field in the critical region. The filled red square is the exact 
    result~\protect{\cite{Sutherland}} for the velocity at $\theta=\pi/4$.
    \label{fig:trimerizedcorrs}
  }
\end{figure}
In order to work out the correlation content of this $S=2$ mode we note that this mode
will only show up in a dynamical correlation function if it is targeted with
an $S=2$ operator. A natural operator is the irreducible quadrupolar part
of the rank two tensor $S^\alpha_iS^\beta_i$. In order to test this conjecture,
we calculate a component of the static spin {\em quadrupolar} structure factor
\be
\mathcal{Q}(k)=  \langle (S^{z})^2(-k) (S^{z})^2(k) \rangle
\ee
and similar for the static spin structure factor
\be
\mathcal{S}(k)=  \langle S^{z}(-k) S^{z}(k) \rangle
\ee
both at the wavevector  $k = 2\pi/3$. 
The results shown in the upper plot of Fig.~\ref{fig:trimerizedcorrs} for system sizes up to $L=18$ display 
nicely that the quadrupolar correlations indeed become more important than the spin-spin 
correlations when one moves beyond the $SU(3)$ symmetric point at $\theta=\pi/4$. 
Although all correlation exponents (dimer, spin, quadrupolar) are equal to
$\eta=2x=4/3$, logarithmic corrections~\cite{ItoiKato} render the quadrupolar correlations clearly  dominant 
in this phase.
Note that our notion of a dominant correlation function in this case is analogous to the
$S=1/2$ antiferromagnetic Heisenberg chain, where both staggered spin and staggered 
dimer correlations have equal correlation exponent $\eta=1$, but the logarithmic corrections 
enhance the spin-spin correlations and then decay more slowly.
Hence, in the same spirit as one argues that the dominant correlations of the Heisenberg chain
are the spin-spin correlations, the dominant correlations of the bilinear-biquadratic spin-one 
chain for $\theta \in (+\pi/4,\pi/2)$ are of spin quadrupolar nature with wavevector $\pm2\pi/3$.

For completeness we show selected parameters of this critical region in the lower panel of
Fig.~\ref{fig:trimerizedcorrs}. The results are in good agreement with field 
theoretical predictions\cite{ItoiKato} ($c=2$ and $x_q=2/3$), exact results\cite{Sutherland}
[$v(\theta{=}\pi/4)=2\pi/3\sqrt{2}]$, and previous numerical work \cite{FathPeriodTripling}. 
The numerical estimate of the scaling dimension $x_q$ is slightly smaller than the
expected value of $2/3$, due to the presence of logarithmic corrections.

To close this section we note that the occurrence of dominant period-three quadrupolar
correlations on a single chain calls for a fully ordered three sublattice structure when the
chains are coupled in an appropriate two-dimensional fashion. The triangular lattice perfectly fulfills 
this prerequisite, and indeed recent works \cite{TsunetsuguArikawa,AMP} reported analytical and
numerical evidence for a three sublattice quadrupolar ordered phase in the region $\theta \in (\pi/4,\pi/2)$.

\section{From the dimerized to the ferromagnetic phase}
\label{sec:nematic}
After having elucidated the nature of the dominant correlations in the extended critical region
beyond the Haldane phase, we now turn to the behavior of the bilinear-biquadratic
chain between the point where the dimerized groundstate is firmly established ($\theta=-\pi/2$)
\cite{DimerizedPhase} and the point where the ferromagnetic state takes over ($\theta=-3\pi/4$).
Let us note that this point $\theta=-3\pi/4$ is special due to a $SU(3)$ symmetry. 
This symmetry has the consequence that the conventional ferromagnetic multiplet is degenerate 
with a fully ordered ferroquadrupolar spin nematic state \cite{BatistaOrtizGubernatis}.

Chubukov~\cite{ChubukovPRB} suggested the existence of an intermediate phase which was
supposed to be completely gapped and to have dominant short-ranged nematic correlations.
Since then several studies \cite{FathNematic,KawashimaPTP,IvanovKolezhuk,LSTpreprint,Rizzi,Porras,Buchta}
 -- most of them numerical -- tried to pin down the existence of an intermediate phase.
Presently it seems most likely that the gapped nematic phase in its original form is not realized in
the single chain model, mainly because the gap data do not show evidence for a closing and reopening.
Nevertheless the spin nematic correlations grow dramatically as one approaches $\theta
\rightarrow -3\pi/4^+$. It could therefore in principle be possible that the dimerized phase gives 
way to a critical spin nematic phase without a gap~\cite{LSTpreprint,Rizzi,Porras}. We will critically 
discuss this possibility in the following.

\subsection{The $S=2$ finite size gap}
\label{sec:chainspintwogap}

We track the evolution of the energy gap to the first magnetic excitation using Density Matrix
Renormalization Group (DMRG) calculations\cite{SRWDmrg} on long open chains of up to
512 spins and retaining up to 1000 states.

\begin{figure}
  \centerline{\includegraphics[width=0.95\linewidth]{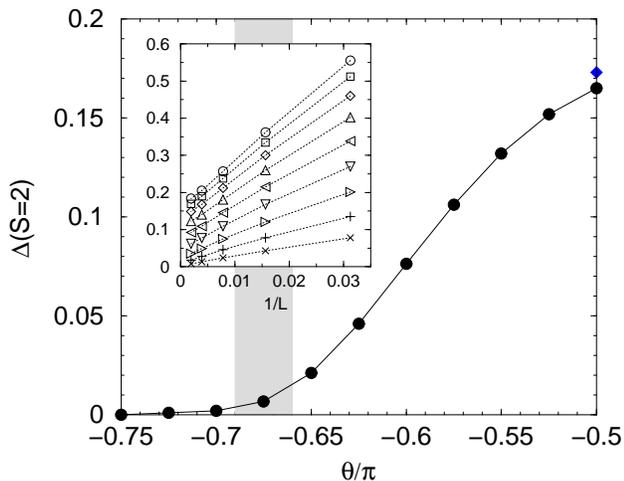}}
  \caption{
    Extrapolated DMRG $S=2$ gaps as a function of $\theta$.
    The gray region denotes the estimated onset of the critical behavior. The
    filled diamond is the exactly known gap result at $\theta=-\pi/2$.
    Inset: finite size extrapolation of the $S=2$ gaps (32 to 512 sites).
    $\theta/\pi$ varies from $-0.5$ to $-0.7$ with decrements of $0.025$ 
    from top to bottom.
    \label{fig:dmrggaptheta}
  }
\end{figure}

A phase transition is signaled by the closing of the gap.
The lowest excited state on open chains carries spin two for $\theta \in (-3\pi/4,-\pi/2]$ 
\cite{FathNematic}. Our results shown in Fig.~\ref{fig:dmrggaptheta} 
are clearly consistent with a finite gap of the $S=2$ excitation for $\theta \gtrsim -0.65 \pi$. 
However, for $\theta \lesssim -0.65 \pi$ the extrapolated gap becomes very small (of the order of $10^{-3}$ for
$\theta = -0.7 \pi$) which could suggest the existence of a phase transition around $-0.67 \pi$
below which the gap is zero. The data shows no evidence for a reopening of the $S=2$ gap in
the interval $\theta \in (-3\pi/4,-0.67\pi]$, at variance with the initial proposal \cite{ChubukovPRB}. 

\begin{figure}
  \centerline{\includegraphics[width=0.95\linewidth]{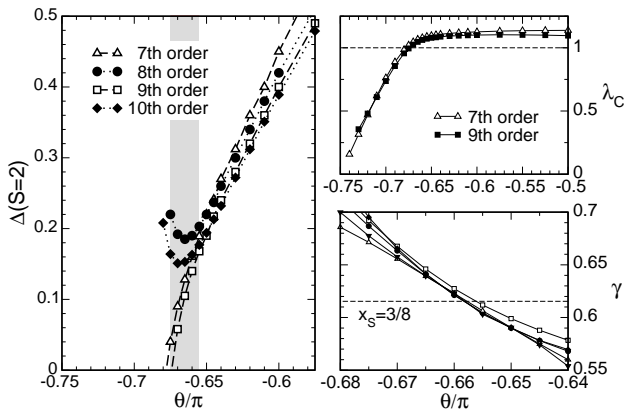}}
  \caption{
    Strong coupling dimer expansion of the $S=2$ gap. 
    Left panel: $S=2$ gap as a function of $\theta$. The gap closes at $\theta_c\approx -0.67 \pi$.
    Upper right panel: Critical value $\lambda_c$ where the gap vanishes. The uniformly coupled 
    chain of interest here corresponds to $\lambda=1$ (dashed line).
    Lower right panel: Critical exponent of the $S=2$ gap opening as a function of the dimerization 
    $(\Delta \sim |\lambda -\lambda_c|^{\gamma})$ calculated by various DLog Pad\'e approximants.
    \label{fig:serieslambda}
  }
\end{figure}

In order to investigate the possible closing of the gap by different means we have calculated strong coupling series expansions\cite{SinghGelfand} of the $S=2$ single particle gap starting from the dimerized limit up to 10th order in the interdimer coupling $\lambda$ for fixed $\theta$.  
A {\it direct} evaluation of the quintuplet gap shows a closing of  the gap around $\theta = -0.67 \pi$ which is illustrated in the left panel of Fig.~\ref{fig:serieslambda}. Using DLog Pad\'e approximants we have determined the critical interdimer coupling $\lambda_c(\theta)$ where the gap closes, see upper right panel of Fig.~\ref{fig:serieslambda}. Consistent with previous estimates by Chubukov \cite{ChubukovPRB} these extrapolations suggest that the gap closes for the uniformly coupled ($\lambda=1$) chain around $\theta \approx -0.67 \pi$.

\begin{figure}
  \centerline{\includegraphics[width=0.95\linewidth]{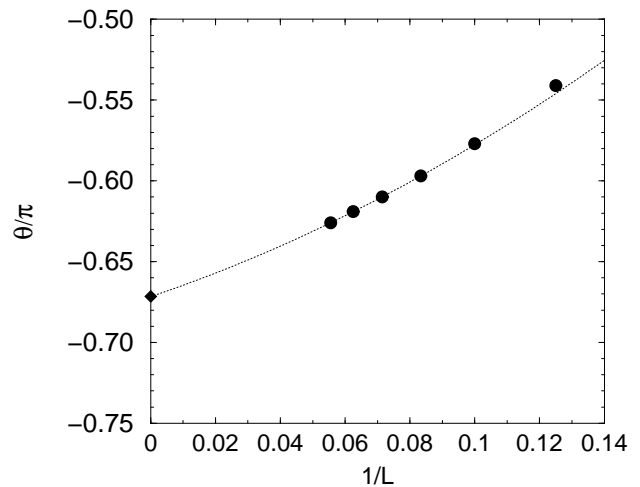}}
  \caption{
    Finite size scaling of the crossing points between the
    singlet gap at momentum $\pi$ and the quintuplet ($S=2$)
    gap at momentum 0. The extrapolation yields a critical 
    value $\theta_c \approx -0.67\pi$ in the thermodynamic limit.
    \label{fig:criticalgaptheta} 
  }
\end{figure}

As a further indication for the possible closure of the gap around $\theta = -0.67 \pi$, we now consider the existence of a generalized BKT phase transition. To determine the critical point we use phenomenological level spectroscopy. We calculate the level crossing $\theta_c^L$  of the lowest singlet excitation at $k=\pi$ with the lowest spin-2 level at $k=0$ for different system sizes $L$ with ED and extrapolate $L \rightarrow \infty$. The results are shown in Fig.~\ref{fig:criticalgaptheta}. The extrapolation is performed with $1/L$ and $1/L^2$ corrections, fitting only the largest four system sizes. The extrapolated critical point is $\theta_c=(-0.67 \pm 0.02) \pi$, in good agreement with estimates obtained by other techniques \cite{Yip,ChubukovPRB}. The coefficient of the $1/L$ correction is small, and fitting without this term yields slightly larger values $\theta_c > -0.67\pi$.

\subsection{Dimerization and quadrupolar correlations}

Dimerization in the phase diagram of the spin-one chain has been firmly established by exact
results obtained for the $\theta = -\pi/2$ point \cite{DimerizedPhase}. It should be noted that 
although the system is dimerized, it is rather poorly described by a simple product wavefunction
of alternating singlets, as can be seen from the small gap and the large correlation length over 
the whole dimerized phase.
The dimerization operator considered here is:
\be
\mathcal{D}(k) \equiv \frac{1}{\sqrt{L}} \sum_{j} e^{i k r_j} (\S_{j}\cdot \S_{j+1})
\ee
with $k=\pi$.
The second kind of correlations expected to be important in this region of the phase diagram 
\cite{PapaUnusualPhases,HaradaKawashimaPRB} are the {\em ferroquadrupolar} spin fluctuations 
\be
\mathcal{Q}(k) \equiv \frac{1}{\sqrt{L}} \sum_{j} e^{i k r_j}  [(S^z_{j})^2-2/3]
\ee
with $k=0$. Note that we considered similar fluctuations with $k=2\pi/3$ in section
\ref{sec:period3}.

\begin{figure}
  \centerline{\includegraphics[width=0.95\linewidth]{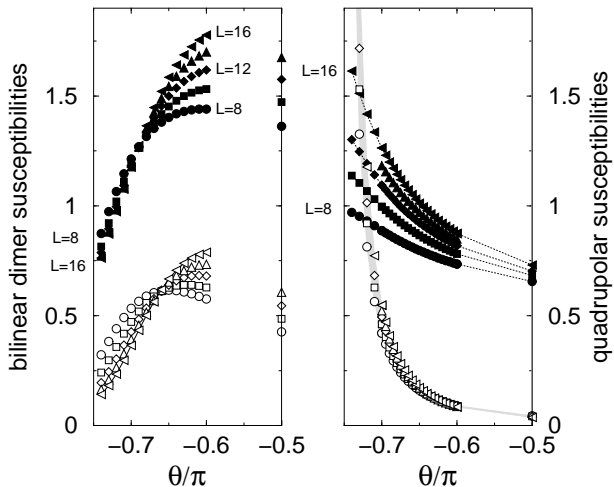}}
  \caption{
    Static and generalized susceptibilities for the bilinear 
    dimerization (left panel) and the ferroquadrupolar correlations (right panel),
    obtained by ED on systems of 8 to 16 sites.
    Solid symbols: Static structure factors.
    Open symbols: Corresponding generalized susceptibilities.
    \label{fig:susceptibilities} 
  }
\end{figure}

Based on these fluctuation operators we have calculated the standard static structure 
factor (SF):
\be
\mathcal{C}^{\mathrm{SF}}(k)= || \mathcal{C}(k)|0\rangle ||^2
\ee
and the generalized nonlinear susceptibility (i.e. the real part of the dynamical 
correlation function at zero frequency)
\be
\mathcal{C}^{\mathrm{GNS}}(k)=\Re \lim_{\eta \rightarrow 0^+} 
\langle0|\mathcal{C}^\dagger(-k)
\frac{1}{H-E_0+i \eta}
\mathcal{C}(k)|0\rangle
\ee
for both kinds of correlations ($\mathcal{C}=\mathcal{D},\mathcal{Q}$). The nonlinear susceptibility
quantifies the perturbative response of the system upon explicitly coupling the symmetry breaking
operator to the Hamiltonian~\cite{NonlinearSusc}.

We present ED results concerning the dimerization in the left panel of Fig.~\ref{fig:susceptibilities}.
The SF and the GNS both diverge as $L\rightarrow \infty$ for $\theta=-\pi/2$, where long range dimer order
is established \cite{DimerizedPhase}.
Similar behavior is found for $\theta \gtrsim -0.67 \pi$.
For $\theta \in (-3\pi/4,-0.67\pi)$ however, we find that the SF and the GNS both {\em decrease} with
system sizes, pointing to a possible absence of spontaneous dimerization in this region.

In the right panel we display the same kind of observables for the $k=0$ ({\em ferroquadrupolar}) 
mode of the quadrupolar correlations. Here the behavior seems different:
while there are only short range correlations deep in the dimerized phase, the ferroquadrupolar 
correlations increase drastically with system sizes for $\theta$ close to $-3\pi/4$.

\begin{figure}
  \centerline{a) $\mathcal{D}(q=\pi,\omega)$}
  \includegraphics[width=0.9\linewidth]{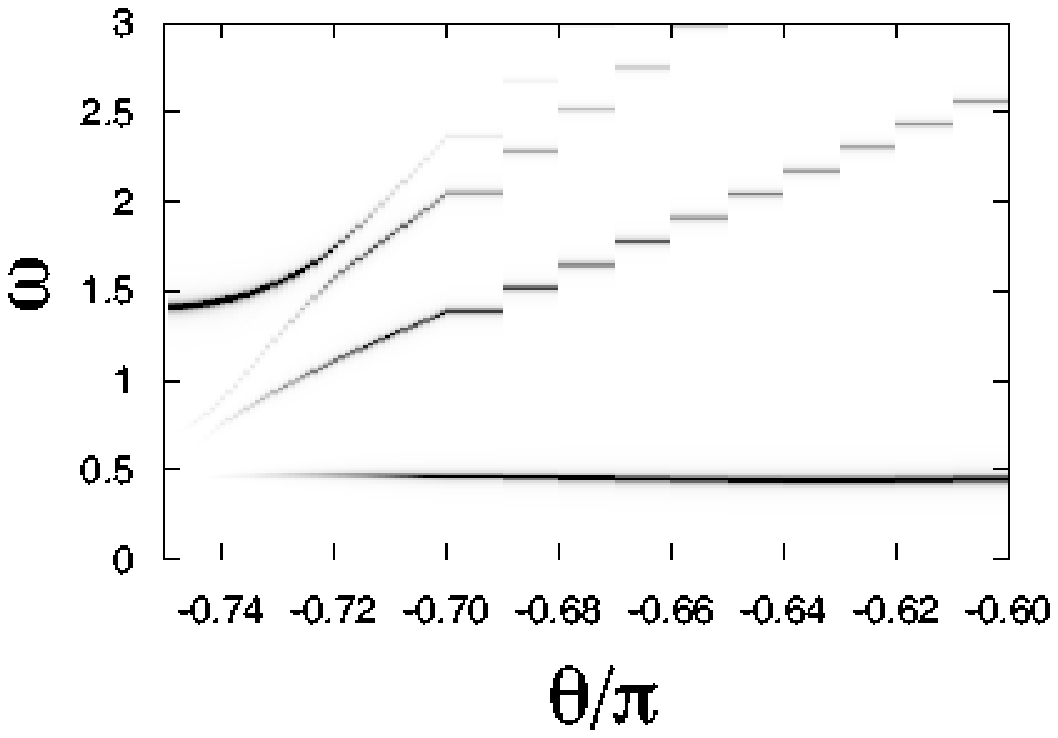} 
 \vspace{3mm}
  $\mbox{}$
  \centerline{b) $\mathcal{Q}(q=0,\omega)$}
  \includegraphics[width=0.9\linewidth]{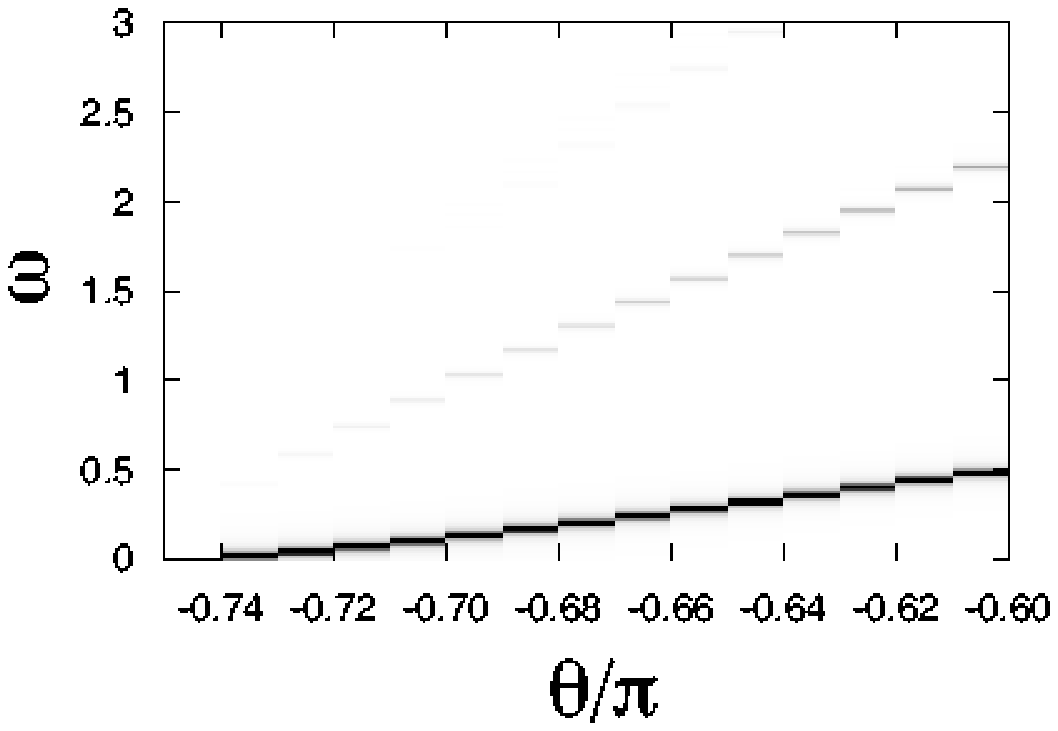} 
  \caption{
    ED dynamical correlation results on a $L=16$ chain. 
    Upper panel:
    dynamical staggered dimer correlation function $\mathcal{D}(q=\pi,\omega)$ 
    plotted for a range of values of $\theta$. There is an important transfer of spectral 
    weight upon lowering $\theta$ from a low energy level -- collapsing to the groundstate as a function
    of system size -- to a level which will converge to the energy $\sqrt{2}$
    at $\theta=-3\pi/4$.
    Lower panel:
    dynamical ferroquadrupolar correlations $\mathcal{Q}(q=0,\omega)$ for $\theta \in
    (0.75,0.60] \pi$.
    \label{fig:dimerdynamics}}
\end{figure}
In order to shed more light on the excitations and their nature close to $\theta=-3\pi/4$ we have 
calculated in addition the {\em dynamical} dimer and quadrupolar structure factor using a continued 
fraction technique. 
This will allow us to track the evolution of the energy of the important levels and their spectral weight. 
The dynamical structure factor is defined as follows:
\be
\mathcal{C}(k,\omega)=- \frac{1}{\pi}\Im 
\lim_{\eta \rightarrow 0^+}  \langle0|\mathcal{C}^\dagger(-k)
\frac{1}{\omega-(H-E_0)+i \eta} \mathcal{C}(k)|0\rangle,
\ee
where we focus again on the dimerization and the ferroquadrupolar correlations. The plots
discussed below show the normalized intensity for a given $\theta$. The overall weight can
be obtained by multiplying with the structure factors displayed in Fig.~\ref{fig:susceptibilities}

The results for the dimerization shown in the upper panel of Fig.~\ref{fig:dimerdynamics}
are very interesting. They show a strong crossover with a redistribution of spectral
weight on the way from $\theta=-\pi/2$ to $-3\pi/4$. While there is a single state (the
lowest energy singlet) exhausting almost all of the $k=\pi$ dimer weight around $\theta=-\pi/2$ 
(as expected in a truly dimer ordered phase), the weight in this state entirely fades out and is transfered
to a higher energy state at $\omega \rightarrow \sqrt{2}$ as $\theta\rightarrow -3\pi/4$. 

The results for the dynamical ferroquadrupolar correlations don't show such a crossover but show 
a steady lowering of the finite size $S=2$ gap upon approaching $-3\pi/4$, accompanied by an 
accumulation of all the weight in that single state. This is a manifestation of the continuity of the
finite size groundstate wavefunction as one approaches the fully ferroquadrupolar {\em ordered}
state at $\theta=-3\pi/4$. We return to the special behavior of the spin-two gap in subsection~\ref{sec:SMA}.

We believe that it is this strong crossover seen in the dynamical dimer correlations which renders numerical 
calculations very difficult. On finite size systems sufficiently close to $\theta=-3\pi/4$ the system looks basically
like a ferroquadrupolar ordered system, even in one dimension, due to the very strong influence of the 
SU(3) point $-3\pi/4$.  As a consequence a crossover scale seems to emerge, which rapidly grows close to
$-3\pi/4$, and one has to go to huge systems in order for the dimerization to win, i.e.~for the lowest 
singlet at $k=\pi$ to collapse onto the groundstate, while the lowest $S=2$ exitation at $k=0$ should remain
gapped. This scenario would then imply that the curve in Fig.~\ref{fig:criticalgaptheta} finally would have to bend 
down for large systems and touch the $y$-axis only at $\theta=-3\pi/4$.

\subsection{Critical phase versus Crossover}
\subsubsection{Parameters of a  hypothetical critical theory}
\begin{figure}
  \includegraphics[width=0.8\linewidth]{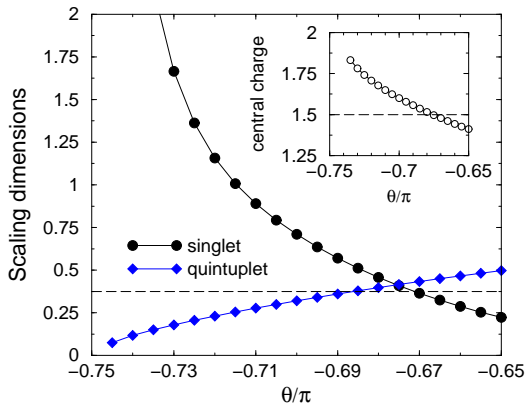} 
  \caption{
   (Color online)
    Hypothetical conformal field theory parameters calculated in ED on chains
    of 8 up to 18 sites. The scaling dimensions of the singlet ($k=\pi$),
    and the quintuplet ($k=0$) field are shown.
    Inset: effective central charge. The value at the boundary of the
    critical region ($\theta \approx -0.67 \pi$) is $c\approx 3/2$.
    \label{fig:scalingdims}
  }
\end{figure}
In order to discriminate between potential conformal theories describing
the phase transition and the extended gapless region, we calculate relevant 
field theory parameters, i.e.~the scaling dimensions (SD) $x$ of several fields
and the central charge $c$ (Fig.~\ref{fig:scalingdims}). The calculation of these
parameters has been performed with ED and relies on finite size scaling properties 
of groundstate and excited states energies \cite{CFTreferences}. Leading logarithmic
corrections have been taken into account. The SD of the $k=\pi$ singlet field ($x_s$) and 
the $k=0$ quintuplet field ($x_q$) are both close to $3/8$ at the onset of criticality near
$\theta\approx -0.67\pi$. $x_s$ significantly increases, while $x_q$ decreases and seems
to approach $0$ as $\theta \rightarrow -3\pi/4$, compatible with quadrupolar long range order
precisely at $\theta=-3\pi/4$ \cite{BatistaOrtizGubernatis}.
Our results for the effective central charge are unexpected. The central charge $c$ does not seem to be
constant throughout the potentially critical region. The smallest value is found at the onset ($c\approx3/2$),
and then seems to increase monotonously. While this behavior can not directly be ruled out on field
theoretical grounds, it is rather uncommon. It remains to be understood whether $c$ is really 
continuously increasing or whether we are facing a crossover phenomenon. 
The hypothetical critical theory at $\theta \approx -0.67\pi$ is however surprisingly well characterized by a 
level two $SU(2)$ Wess-Zumino-Witten model given that both the scaling dimensions ($3/8$) and the 
central charge ($3/2$) are in agreement with such a theory.  
Some more support for this claim comes from the indirect calculation of $x_s$ within our
series expansions. There the critical exponent $\gamma$ of the $S=2$ gap is related to the scaling 
dimension $x_s$ by $\gamma = 1/(2-x_s)$. In the lower right panel of Fig.~\ref{fig:serieslambda} we show
the critical exponent calculated by DLog Pad\'e approximants to the 10th order series. In the  vicinity 
of $\theta \approx -0.67\pi$ the various approximants reveal only small spreading. Within the precision 
of a 10th order calculation the exponents comply with a scaling dimension of  $x_s = 3/8$ (dashed line).

\subsubsection{Single Mode approximation}
\label{sec:SMA}

In the following we present an argument based on the single mode 
approximation (SMA) which aims at explaining the anomalously 
strong suppression of the $S=2$ gap upon approaching $-3\pi/4$.

In the single mode approximation one starts by constructing a trial
state upon the application of a structure factor operator on the groundstate:
\be
|\mathcal{O}(k)\rangle=\mathcal{O}(k) |0\rangle\,,
\ee
where 
\be
\mathcal{O}(k) = \frac{1}{\sqrt{L}}\sum_{j} e^{i k r_j} \mathcal{O}_j\,,
\ee
and $\mathcal{O}_j$ is an operator acting on site $j$.
One can now relate the variational energy of the state $|\mathcal{O}(k)\rangle$ to a groundstate 
expectation value:
\be
\omega_k=\frac{1}{2}\frac{\langle 0|[\mathcal{O}^\dagger(-k),[H,\mathcal{O}(k)]] |0\rangle}
{\langle 0|\mathcal{O}^\dagger(-k)\mathcal{O}(k) |0\rangle}\,.
\ee
Note that the energy $\omega_k$ is a strict upper bound on the gap in the momentum $k$ sector.
The power of the SMA comes from the fact that it can be used to prove the absence of a gap 
under certain conditions.
There are two prototypical cases for the absence of a gap:
i) The denominator diverges as a function of systems size, while the numerator diverges more slowly or 
stays finite. This is the conventional situation for systems which are critical or exhibit spontaneous 
symmetry breaking in the thermodynamic limit. In this case the diverging static structure factors drives 
the gap to zero, but only for the infinite system. 
ii) The numerator vanishes, while the denominator does not vanish. A sufficient condition for the numerator
to vanish is if the structure factor operator $\mathcal{O}(k)$ commutes with the Hamiltonian.

If we now choose for the operator $\mathcal{O}(k)$ the spin quadrupolar structure factor at zero momentum:
i.e. $\mathcal{Q}(0) = 1/\sqrt{L} \sum_{i}\left[(S^z)_i^2-2/3\right]$ we realize scenario ii) at the special $SU(3)$
point $\theta=-3\pi/4$. For this specific value $\mathcal{Q}(0)$ commutes with the Hamiltonian and therefore the 
numerator is zero. Numerically we find that the denominator is different from zero. This means that there is at 
least a level with $S=2$ which is degenerate with the groundstate already on finite size samples~\cite{FathNematic}.
The reason which makes the numerical analysis for $\theta > -3\pi /4$ so difficult is that there is  a continuity in 
the wavefunction as a function of $\theta$ when approaching the $SU(3)$ point from above. On one hand
we know that the commutator $[\mathcal{Q}(0),H(\theta)]$ will vanish continuously as we approach $-3\pi/4^+$,
on the other hand we find numerically that the denominator (the ferroquadrupolar structure factor) grows rapidly 
as a function of the accessible system sizes, c.f. full symbols in the right panel of Fig~\ref{fig:susceptibilities}. 
So these two factors cooperate in giving an anomalously small SMA energy $\omega_k$, which is an upper 
bound to the measured gap. 

\section{Two coupled chains}
\label{sec:twochains}
\begin{figure}
  \includegraphics[width=0.95\linewidth]{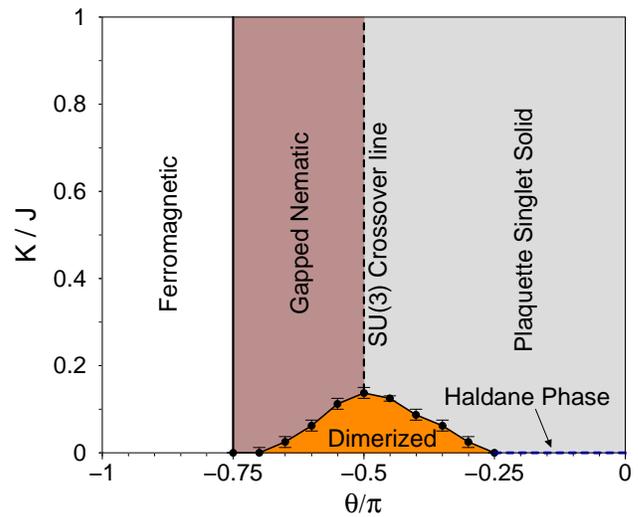}
  \caption{
    (Color online) Phase diagram of two coupled bilinear-biquadratic
    $S=1$ chains. The dimerization along the chains is rapidly suppressed by
    a finite interchain coupling. The dominating phase is a unique gapped 
    phase which is adiabatically connected to the Haldane phase of two isolated
    Heisenberg chains.
    \label{fig:twochains}}
\end{figure}
We have seen considerable difficulty in the preceding section to actually decide
whether the dimerization vanishes before reaching $\theta = -3\pi/4$.
One potential way of circumventing the spontaneous dimerization of a single chain
is to consider two coupled chains in a ladder geometry. It seems rather natural to 
assume that the dimerization of a single chain disappears once the coupling
on the rung is suffiently strong. We will substantiate this claim using numerical
simulations below. The ladder model is also a first step towards a realistic 
setup of spin-1 bosonic atoms in optical lattices, where a finite interchain
coupling can easily be generated.
The ladder Hamiltonian is given as follows:
\bea
H&=& J \sum_{i,n} \cos \theta \left(\S_{i,n}\cdot\S_{i+1,n} \right) 
+ \sin \theta \ \left(\S_{i,n} \cdot\S_{i+1,n} \right)^2 \nonumber\\
&+ K & \sum_{i} \cos \theta \left(\S_{i,1}\cdot\S_{i,2} \right) 
+ \sin \theta \ \left(\S_{i,1} \cdot\S_{i,2} \right)^2,
\label{eqn:biquadLadderHamiltonian}
\eea
where $\S_{i,n}$ denotes a spin-one operator at position $i$ on chain $n \in \{1,2\}$.
In the following we choose $K,J>0$ and vary $\theta$ as well as $K/J>0$, where the 
limit $K/J=0$ corresponds to decoupled chains and $J/K=0$ to decoupled rung dimers.

For $\theta=0$ the Hamiltonian (\ref{eqn:biquadLadderHamiltonian}) describes a 
conventional $S=1$ Heisenberg spin ladder \cite{LadderPapers,TodoLadder}.
Interestingly, for this special case Todo {\it et al.}~\cite{TodoLadder}
have shown numerically that the finite gap of the single chain does {\em not} close for
any $K/J$, i.e two points on the $K/J$ axis can be reached without any second 
order quantum phase transition in between. 
They also generalized the string order parameter of a single chain to a more complicated 
nonlocal operator which is non-zero for any finite value of $K/J$, therefore showing that
this phase is also topologically ordered~\cite{TodoLadder}.
In the following we will show that this gapped spin liquid state at
$\theta=0$ extends deeply into the region $\theta\le0$, and for suffiently large $K/J$ even to $-3\pi/4^+$.
While the groundstate is always protected by a finite gap, the nature of the lowest excitation
changes as a function of $\theta$. For $\theta \in [-\pi/2,0]$ the lowest excitation carries
$S=1$, while for $\theta \in [-3\pi/4,-\pi/2]$ it carries $S=2$. In the absence of dimerization
such a state on the ladder where the lowest excitation is of ferroquadrupolar nature,
is very close in spirit to the long-sought gapped nematic phase initially 
proposed by Chubukov~\cite{ChubukovPRB} for the single chain. The phase diagram of two coupled chains
with $\theta \in [-\pi,0]$ and $K/J \in [0,1]$ is shown in Fig.~\ref{fig:twochains}. We expect
the extended gapped phase to continue up to the rung dimer limit $J/K=0$, similar to the results
for $\theta=0$~\cite{LadderPapers,TodoLadder}.

\begin{figure}
  \includegraphics[width=0.95\linewidth]{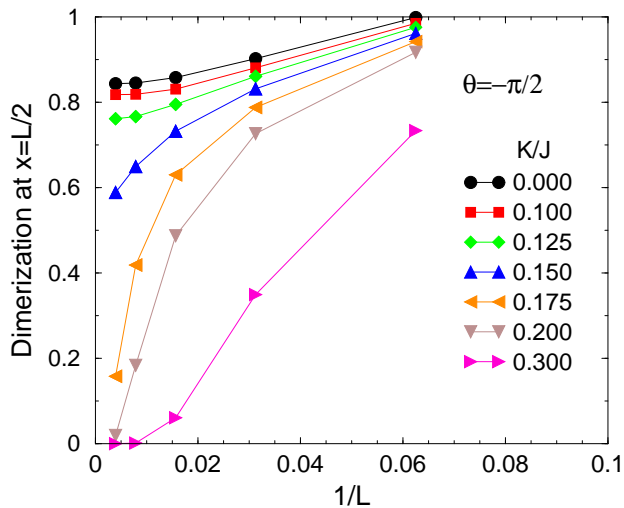}
  \caption{
    (Color online)
    Finite size scaling of the dimerization measured in the 
    middle of a ladder as a function of $K/J$.
    It can be clearly seen that that the dimerization extrapolates
    to a finite value for $K/J\le 0.125$, while it tends to zero
    for $K/J\ge 0.15$.
    \label{fig:ladderdimerization}}
\end{figure}
Let us now describe the simulation results which lead to the phase diagram 
presented in Fig.~\ref{fig:twochains}.
First we discuss the behavior of the dimerization of the single chain upon coupling 
two chains. To this end we performed DMRG calculations on ladder systems of sizes
up to $2\times 256$ sites.
We have determined the boundary of the dimerized phase using finite size extrapolations of the
remnant dimerization in the middle of a chain~\cite{WhiteJ1J2}, 
assuming a columnar dimer arrangement on the two chains \footnote{
  This seems to be a reasonable assumption, given the fact that 
  $SU(N)$ spin models on the square lattice dimerize this way for $N>4$~\cite{HaradaSUN},
  while our model at $\theta=-\pi/2$ corresponds to $N=3$.
}.
An example for such finite size data is shown for $\theta=-\pi/2$ in 
Fig.~\ref{fig:ladderdimerization}. We expect the dimerization in the middle of the chain
to converge to a finite value for $1/L\rightarrow 0$ in a dimerized phase, and to
exhibit an exponential drop upon reaching the correlation length of a disordered phase.
These two distinct behaviors can be seen in Fig.~\ref{fig:ladderdimerization} for the
values $K/J=0,0.1,0,125$ and $K/J=0.15,0.175,0.2,0.3$ respectively. This leads us
to the conclusion of a critical ratio $(K/J)_c=0.1375\pm 0.0125$ at $\theta=-\pi/2$,
below which the ladder is still dimerized. The evolution of the values of $(K/J)_c$ with
$\theta$ is shown in Fig.~\ref{fig:twochains}. The dimerized phase is most stable close
to $\theta=-\pi/2$, and seems to vanish linearly upon approaching the BT point at $-\pi/4$,
similar to the case of explicit dimerization for a single chain~\cite{KitazawaBT}.
On the other side however the dimerization boundary drops rapidly and becomes very small
once $\theta\lesssim -0.7\pi$. Similar to the single chain case it is very difficult to decide whether
the dimerized phase ceases to exist before $-3\pi/4$ or not. In the ladder case however
this implies that for a finite -- but very small -- interchain coupling the spontaneous dimerization 
(if any) of a single chain vanishes. 

\begin{figure}
  \includegraphics[width=0.95\linewidth]{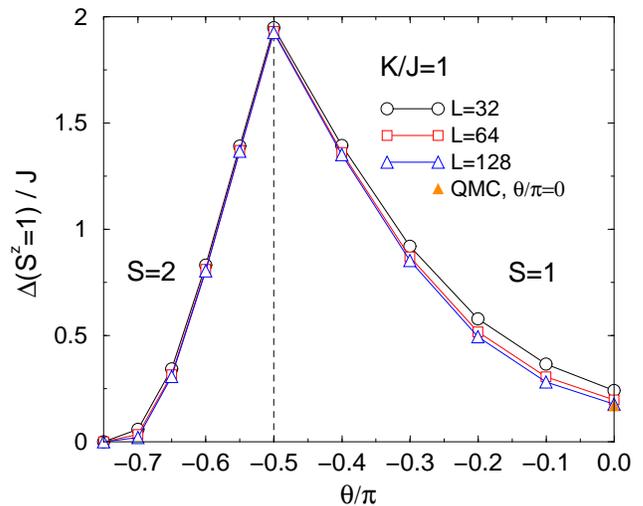}
  \caption{
    (Color online)
    Evolution of the spin gap for constant $K/J=1$ as a function
    of $\theta \in [-3\pi/4,0]$. The nature of the gap changes
    at the $SU(3)$ point $\theta=-\pi/2$, from a $S{=}1$ state for $\theta>-\pi/2$
    to a a $S{=}2$ state for $\theta<-\pi/2$. The QMC value of the
    gap at $\theta=0$ is from Ref.~[\protect{\onlinecite{TodoLadder}}].
    \label{fig:laddergap}}
\end{figure}
Next we investigate the behavior of the spin gap in the non-dimerized phase. We performed
DMRG calculations for a fixed value of $K/J=1$. The results for the gap to the lowest $S^z=1$
state are plotted in Fig.~\ref{fig:laddergap} for system sizes $2\times L$ with $L=32,64,128$.
Starting at $\theta=0$ where we find good agreement with the high-precision QMC gap from 
Ref.~[\onlinecite{TodoLadder}], the gap grows monotonously until reaching the bipartite
$SU(3)$ point at $\theta=-\pi/2$. There we observe a level crossing, and the gap decreases
for $\theta<-\pi/2$. Note that at this special point the total spin quantum number of the lowest
level changes. While it is a $S=1$ state for  $\theta>-\pi/2$, the lowest levels carries
$S=2$ for $\theta<-\pi/2$.
Similar to the single chain case discussed in Sec.~\ref{sec:chainspintwogap},
we detect a very small gap for $\theta$ close to $-3\pi/4$. This anomalously small $S=2$
gap can again be understood using a single mode approximation argument, along the lines 
of Sec.~\ref{sec:SMA}, the extension of the argument to the ladder case being straightforward.
We take this dichotomy of the gap as a first indication that the $SU(3)$ line ($\theta=-\pi/2$)
constitutes a crossover line inside the extended gapped phase, where the spin-spin 
correlations dominate for $\theta \gtrsim -\pi/2$, while we expect dominant short-ranged
spin nematic ferroquadrupolar correlations for $\theta\lesssim -\pi/2$. Our gap results 
give evidence that the groundstate for $\theta$ close to $-3\pi/4$ adiabatically connects
to the "Plaquette Singlet Solid" phase, which is itself connected to the Haldane phase of the
isolated chains. This would also imply that the the whole phase is characterized by a finite 
nonlocal order parameter. It would be interesting to investigate this order in a future work.

\section{Conclusions}
To summarize, we have shown that spin nematic correlations in the form of quadrupolar 
correlations play an important role in the phase diagram of single or coupled $S=1$ 
bilinear-biquadratic chains. 

We first addressed the long-standing question of the nature of the dominant correlations in
the gapless period-three phase, where we uncovered the spin quadrupolar correlations at 
$k=\pm 2\pi/3$ as the leading ones. We critically discussed the phase diagram close
to the $SU(3)$ point at $\theta=-3\pi/4$ and by assembling several numerical and analytical
results concluded that an unconventional crossover phenomenon is at the heart of the considerable
difficulty in settling the issue on the existence or absence of the gapped nematic phase put 
forward by Chubukov~\cite{ChubukovPRB}. 
Finally we studied two coupled bilinear-biquadratic chains in a ladder geometry and found 
an extended gapped phase which is surprising in two respects: first it realizes a variant
of Chubukov's nematic phase, i.e. a gapped, non-dimerized phase with dominant spin-nematic
correlations for $\theta\lesssim -\pi/2$, and second, this phase is adiabatically connected to
the standard $S=1$ Heisenberg ($\theta=0$) ladder, and therefore also connected to the
Haldane phase of single  chains, as shown by Todo {\it et al.}~\cite{TodoLadder}. 

\acknowledgments
We would like to thank
F.~Alet,
F.D.M.~Haldane,
K.~Harada,
N.~Laflorencie,
P.~Lecheminant,
F.~Mila,
D.~Roggenkamp,
M.~Troyer
and 
S.~Wessel for valuable discussions.
We are grateful to S. Todo for providing us his ladder QMC gap data.
The authors acknowledge support from the Swiss National Science Foundation.
Part of the numerical calculations have been performed on the ``Pl\'eiades'' 
cluster at EPFL.

\end{document}